\documentclass[12pt]{iopart}

\expandafter\let\csname equation*\endcsname\relax
\expandafter\let\csname endequation*\endcsname\relax

\usepackage{amsmath}
\usepackage{bm}
\usepackage{graphicx}
\usepackage[hidelinks]{hyperref}
\usepackage{xcolor}
\usepackage[notrig]{physics}
\usepackage{subfigure}

\newcommand{\calO}{O}
\newcommand{\s}{\sigma}
\newcommand{\vs}{\bm{\sigma}}
\newcommand{\vxi}{\bm{\xi}}

\begin{document}

\title{Sparse Autoregressive Neural Networks for Classical Spin Systems}
\author{Indaco Biazzo$^{1,*}$, Dian Wu$^1$, Giuseppe Carleo$^1$}
\address{$^1$ Institute of Physics, \'Ecole Polytechnique F\'ed\'erale de Lausanne (EPFL), CH-1015 Lausanne, Switzerland}
\ead{indaco.biazzo@gmail.com}

\vspace{10pt}

\begin{abstract}
Efficient sampling and approximation of Boltzmann distributions involving large sets of binary variables, or spins, are pivotal in diverse scientific fields even beyond physics. Recent advances in generative neural networks have significantly impacted this domain. However, these neural networks are often treated as black boxes, with architectures primarily influenced by data-driven problems in computational science.
Addressing this gap, we introduce a novel autoregressive neural network architecture named TwoBo, specifically designed for sparse two-body interacting spin systems. We directly incorporate the Boltzmann distribution into its architecture and parameters, resulting in enhanced convergence speed, superior free energy accuracy, and reduced trainable parameters.
We perform numerical experiments on disordered, frustrated systems with more than $1000$ spins on grids and random graphs, and demonstrate its advantages compared to previous autoregressive and recurrent architectures.
Our findings validate a physically informed approach and suggest potential extensions to multivalued variables and many-body interaction systems, paving the way for broader applications in scientific research.
\end{abstract}

\section*{Introduction}

This paper addresses the challenge of efficiently sampling and approximating the Boltzmann distribution in systems characterized by two-body interacting binary variables, or spins.
The interactions among the variables are typically defined on sparse graphs, such as lattice systems pivotal in material science, and random graphs that model a plethora of complex socio-technological systems~\cite{doi:10.1126/science.286.5439.509, RevModPhys.74.47}.
The complexity of the system's Boltzmann distribution, and thus its underlying physics, significantly depends on the couplings and external field values. For instance, disordered or random coupling values lead to complex and frustrated systems that could reduce the efficiency of sampling algorithms such as Monte Carlo approaches. This complexity finds relevance in various areas, including spin glasses~\cite{10.1142/0271}, optimization on random graphs~\cite{mezard2009information}, and statistical inference problems~\cite{doi:10.1080/00018732.2016.1211393}.

The accurate approximation and efficient sampling of the Boltzmann distribution is crucial to solving problems in various fields, ranging from biology~\cite{Cocco_2018} to computer science~\cite{doi:10.1126/science.1073287}. The recent advent of deep generative models, particularly autoregressive neural networks (ARNNs), has marked a significant advancement in fields such as image and language generation~\cite{NEURIPS2020_1457c0d6}. A pivotal study in 2019 introduced the use of ARNNs for Boltzmann distribution sampling~\cite{PhysRevLett.122.080602}, leading to a proliferation of research in both classical~\cite{10.1103/physreve.101.023304, PhysRevE.101.053312, PhysRevE.103.012103, PhysRevResearch.3.L042024, hibat-allah_variational_2021} and quantum physics~\cite{PhysRevLett.124.020503, 10.1103/physrevlett.128.090501, PhysRevA.102.062413, PhysRevResearch.2.023358, Liu_2021, Barrett2022, Cha_2022, PhysRevResearch.5.L032001}, as well as statistical inference~\cite{Biazzo2022} and optimization problems~\cite{khandoker2023supplementing}.

The ARNN architectures used are typically adapted from those developed for data-driven problems in computer science, and prior knowledge on theoretical properties of the physical system has rarely been utilized to customize the ARNN architecture, or only been exploited for specific physical systems~\cite{BIALAS2022108502, Biazzo2022}.
Recently, a direct mapping was established between the Boltzmann distribution and the corresponding ARNN architecture~\cite{biazzo2023autoregressive}. The derived general architecture was not directly usable, but thanks to the analytic derivation, two specific architectures for two well-known fully connected spin models, the Curie--Weiss model and the Sherringhton--Kirkpatrick model, were derived.

The contribution of this work is a novel ARNN architecture tailored for general two-body interacting systems, which exploits the knowledge of the Hamiltonian and the Boltzmann distribution.
The proposed architecture has some of its parameters precomputed from the Hamiltonian couplings and takes advantage of the usual sparsity of the interactions among spins. It has fewer trainable parameters and a faster convergence speed than conventional ARNN architectures while maintaining a high expressivity.
This universal method can be applied across two-body interacting systems, offering a more efficient sampling approach than standard ARNN architectures and opening the way to richer, physically informed neural network architectures.

\section*{Methods}

Consider a system of $N$ spins characterized by a Hamiltonian $H$. These spins, each denoted as $\s_i$, can take values in $\pm 1$. Let $\vs = \{\s_1, \s_2, \ldots, \s_N\}$ represent the set of all spins. The Hamiltonian is given by $H(\vs) = -\sum_{\langle i, j \rangle} J_{i j} \s_i \s_j - \sum_i h_i \s_i$, where the summation $\sum_{\langle i, j \rangle}$ extends to all interacting pairs of spins. The term $J_{i j}$ represents the interaction strengths between the spins, and $h_i$ is the external magnetic field acting on each spin.

We approximate the Boltzmann distribution $P(\vs)$ using an ARNN whose probability distribution is decomposed in the autoregressive form $Q(\vs) = \prod_i Q_i(\s_i \mid \vs_{< i})$, where $\vs_{< i} = \{\s_1, \s_2, \ldots, \s_{i - 1}\}$ denotes the set of variables with indices less than $i$. If we can write a probability distribution in this form, then the ancestral sampling procedure, where each variable $\s_i$ is sampled in the subsequent order according to its conditional probability $Q_i(\s_i \mid \vs_{< i})$ given the previous variables, produces guaranteed independent samples.

We parameterize each conditional probability with a neural network. Without limitation to particular interacting systems, variable sharing schemes, or recurrent structures, one may use feedforward neural networks to parameterize the conditional probabilities, with masked dense layers to ensure the autoregressive property. This architecture is known as masked autoencoder for distribution estimation (MADE), which was first introduced in~\cite{pmlr-v37-germain15} and used in physics in~\cite{PhysRevLett.122.080602}. However, both its number of parameters and the computation time of each layer scale by $\calO(N^2)$ with the system size $N$, limiting its applicability to small systems.

\begin{figure}[htb]
\centering
\subfigure[]{\includegraphics[width=0.58\linewidth]{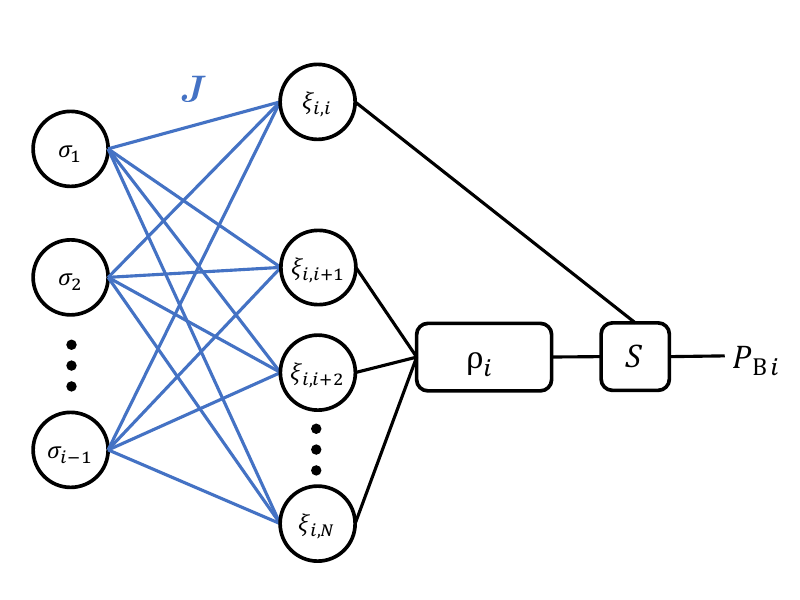}}
\hfill
\subfigure[]{\includegraphics[width=0.41\linewidth]{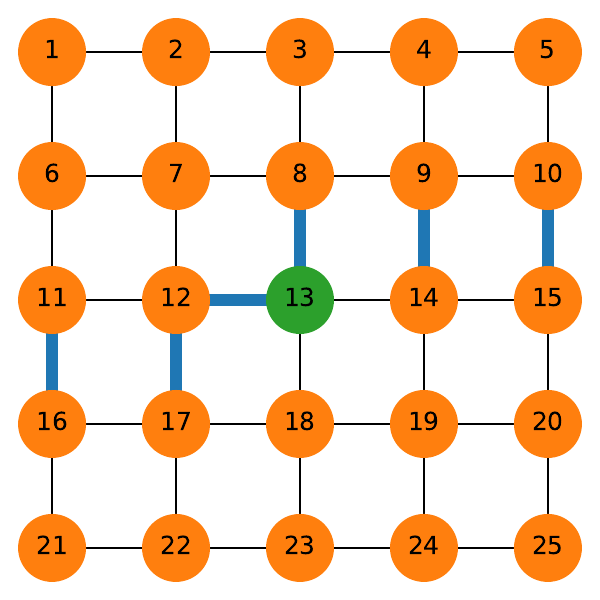}}
\caption{
(a) Sketch of the TwoBo neural network architecture to compute a conditional probability $P_{\text{B} i}(\s_i = +1 \mid \vs_{< i})$ in the autoregressively decomposed Boltzmann distribution.
The parameters of the first layer are taken from the coupling matrix $\bm{J}$.
The function $\rho_i$ is parameterized as a feedforward neural network, whose inputs are the variables $\vxi_i$ after the first layer.
The last layer $S$ is a sigmoid function.
(b) Calculating the conditional probability $P_{\text{B} i}$ with $i = 13$ on a 2D grid of $N = 25$. The edges between the spins $\vs_{< i}$ and $\vs_{\ge i}$ are highlighted in blue, and only those edges are used when computing $\xi_{i l}$. Among the variables $\{\xi_{i l} \mid l > i\}$, only those with $l < i + L$ are kept as inputs to $\rho_i$, and the others are ensured to be zero.
}
\label{fig:arch_grid}
\end{figure}

In the following, we will show how the knowledge of the Hamiltonian and the Boltzmann distribution leads to systematically reducing both the number of trainable parameters and the computation time.
Consider the Boltzmann distribution $P_\text{B}(\vs) = \frac{\rme^{-\beta H(\vs)}}{Z}$, where $\beta$ is the inverse temperature and $Z = \sum_{\vs} \rme^{-\beta H(\vs)}$ is the normalization factor. Following the derivation presented in~\cite{biazzo2023autoregressive}, we can write:
\begin{equation}
P_{\text{B} i}(\s_i \mid \vs_{< i})
= \frac{\sum_{\vs_{> i}} P_\text{B}(\vs)}{\sum_{\vs_{> i - 1}} P_\text{B}(\vs)}
= \frac{\sum_{\vs_{> i}} \rme^{-\beta H(\vs)}}{\sum_{\vs_{> i - 1}} \rme^{-\beta H(\vs)}}
= \frac{f(\s_i, \vs_{< i})}{\sum_{\s_i} f(\s_i, \vs_{< i})},
\end{equation}
where we defined $f(\s_i, \vs_{< i}) = \sum_{\vs_{> i}} \rme^{-\beta H(\vs)}$. The positive conditional probability of the variable $\s_i$ can be written as:
\begin{equation}
P_{\text{B} i}(\s_i = +1 \mid \vs_{< i})
= \frac{f(\s_i = +1, \vs_{< i})}{f(\s_i = +1, \vs_{< i}) + f(\s_i = -1, \vs_{< i})}
= S\left( \log \frac{f(\s_i = +1, \vs_{<i})}{f(\s_i = -1, \vs_{< i})} \right),
\end{equation}
where we impose the sigmoid function $S(x) = \frac{1}{1 + \rme^{-x}}$ as the last layer of the neural network. This ensures that the network output is constrained to lie between zero and one. Substituting the definition of the Hamiltonian and the Boltzmann distribution, we obtain the following:
\begin{equation}
P_{\text{B} i}(\s_i = +1 \mid \vs_{<i}) = S\left( 2 \beta \sum_{s < i} J_{s i} \s_s + 2 \beta h_i + \log \frac{\rho_i^+(\vs_{< i})}{\rho_i^-(\vs_{< i})} \right),
\label{eq:cond_prob}
\end{equation}
where:
\begin{equation}
\rho_i^{\pm}(\vs_{< i}) = \sum_{\vs_{> i}} \exp \beta \left( \sum_{l > i} \s_l \left( \pm J_{i l} + \sum_{s < i} J_{s l} \s_s + h_l \right) + \sum_{l, l' > i} J_{l l'} \s_l \s_{l'} \right).
\label{eq:rho_ghann}
\end{equation}
Observing $\rho_i^{\pm}$ as a sequence of operations on the input variables $\vs_{<i}$, the first operation is a linear transformation, which leads us to define a linear layer with the outputs:
\begin{equation}
\xi_{i l} = \sum_{s < i} J_{s l} \s_s \quad \text{with} \quad l \ge i.
\label{eq:s_il_first}
\end{equation}
Now Eq.~\eqref{eq:cond_prob} can be written as:
\begin{equation}
P_{\text{B} i}(\s_i = +1 \mid \vs_{< i}) = S\left( 2 \beta \xi_{i i} + 2 \beta h_i + \rho_i(\vxi_i) \right),
\label{eq:twobo_arch}
\end{equation}
where we defined $\rho_i(\vxi_i) = \log \frac{\rho_i^+(\vxi_i)}{\rho_i^-(\vxi_i)}$ and $\vxi_i = \{\xi_{i, i + 1}, \xi_{i, i + 2}, \ldots, \xi_{i, N}\}$. The variables $\vxi_i$ can be explicitly computed given the input spins $\vs_{< i}$ and the Hamiltonian parameters $\bm{J}$, which means that the first layer of our neural network does not involve any trainable parameter.
After the first layer, the term $2 \beta \xi_{i i}$ appears as an input to the last layer $S$, which can be considered as a skip connection~\cite{He_2016_CVPR}.
As shown in Eq.~\eqref{eq:rho_ghann}, the exact computation of the function $\rho_i$ involves a sum over all configurations of $\vs_{> i}$, which scales exponentially with the system size.
The entire network architecture to compute Eq.~\eqref{eq:twobo_arch} is sketched in Fig.~\ref{fig:arch_grid}~(a).

In~\cite{biazzo2023autoregressive}, the functions $\rho_i$ were approximated with specific architectures derived for the fully connected Curie--Weiss and Sherrington--Kirkpatrick models. In the present work, we propose a more general approach to approximate $\rho_i$ using feedforward neural networks, whose inputs are the variables $\vxi_i$, and to take advantage of the sparsity of interactions among the spins.
In the following, we refer to this architecture as TwoBo (two-body interactions).

When computing a conditional probability $P_{\text{B} i}$ with a fixed $i$, the computation time of the first layer, Eq.~\eqref{eq:s_il_first}, scales linearly in the number of non-zero elements in $\bm{J}$. For fully connected interactions, it scales as $\calO(N^2)$ with the system size $N$; whereas for sparse interactions, it is linear in $N$.
Moreover, in the case of sparse interactions, many of the variables in $\vxi_i$ are ensured to be zero. As seen in Eq.~\eqref{eq:s_il_first}, if the spin $\s_l$ does not interact with all spins $\vs_{< i}$, then $\xi_{i l}$ is zero and can be discarded in the parameterization of the function $\rho_i$.
Consider, for instance, a 2D grid of side length $L$, with spins $\vs$ ordered as in Fig.~\ref{fig:arch_grid}~(b). The variables $\xi_{i l}$ are non-zero only when $i < l < i + L$, so the number of variables in $\vxi_i$ is reduced from $\calO(N) = \calO(L^2)$ to $\calO(L)$.
Furthermore, since the Hamiltonian is locally interacting, each remaining $\xi_{i l}$ only depends on a fixed number of spins, which does not scale with $N$, so the computation time of the first layer is only $\calO(L)$.
When we compute all conditional probabilities $\{P_{\text{B} i} \mid 1 \le i \le N\}$ in a batch, the computation time of the first layer is $\calO(L N) = \calO(L^3)$, which is polynomially lower than $\calO(L^4)$ of MADE.
After the first layer, we use a lightweight neural network for $\rho_i$, whose computational complexity does not exceed $\calO(L^3)$. The empirical scaling of the number of trainable parameters is shown in the Supplementary Material~\ref{append:param}.
Similar polynomial reduction of trainable parameters is also valid with other local interactions, such as second and third nearest neighbors, as long as the interaction range is much less than the system size.

When training the neural network in a variational setting, we denote it as $Q^\theta(\vs)$, where $\theta$ is the set of trainable parameters. The parameters are trained to minimize the Kullback--Leibler divergence $D_\text{KL}(Q^\theta \mid\mid P) = \sum_{\vs} Q^\theta(\vs) \log \frac{Q^\theta(\vs)}{P(\vs)}$ between the variational and the target distributions. When the target is the Boltzmann distribution $P_\text{B}$, it is equivalent to minimizing the variational free energy $F^\theta = \sum_{\vs} Q^\theta(\vs) \left( \frac{1}{\beta} \log Q^\theta(\vs) + H(\vs) \right)$, except for a constant term. Its exact computation involves a summation over the spin configurations that grows exponentially with the system size. In~\cite{PhysRevLett.122.080602}, it was proposed to estimate the variational free energy and its gradients with respect to $\theta$ by a finite subset of configurations sampled from the ARNN thanks to the ancestral sampling procedure.

In this work, to alleviate the mode collapse problem, an annealing procedure is used in the training, starting from a low value of $\beta = 0.05$ (high temperature) and changing it in steps of $0.05$ until $\beta = 3$. In each temperature step, a fixed number of optimization steps are executed, then the trained neural network is used as the initialization at the next temperature. In the case of TwoBo, the $\beta$ used in the skip connection coefficient in Eq.~\eqref{eq:twobo_arch} is also updated accordingly, and its effect is shown in the Supplementary Material~\ref{append:beta}.

\section*{Results}

We test the performance of TwoBo in approximating the Boltzmann distribution of sparse and disordered interacting spin systems.
We first choose Edward--Anderson (EA) models~\cite{edwards1975theory} on 2D and 3D grids with periodic boundary conditions as target distributions, which can demonstrate the rich physics of the spin glass phase~\cite{10.1142/0271, picco2001chaotic, stein2013spin}. In particular, the 3D EA model falls into the NP-complete class of computational complexity~\cite{cipra2000ising}, and has been a widely recognized testing ground for algorithms aiming to solve computationally difficult optimization problems~\cite{amey2018analysis, cirauqui2024population}.
Moreover, we consider random regular graphs (RRG) as another target, which naturally arises in modeling long-range interactions and cooperative behaviors in several domains, such as biological and techno-social systems~\cite{mezard2001bethe, RevModPhys.74.47,dommers2017metastability}.

The couplings $J_{i j}$ are randomly sampled to be either $+1$ or $-1$, and the external fields $h_i$ are set to zero for simplicity. Each result we report in this paper is averaged from $10$ random instances, and for each instance we train a neural network, respectively.
When generating the RRGs, we fix the degree $d = 3$.

Although in general we can use a deep feedforward neural network to parameterize each function $\rho_i$ in Eq.~\eqref{eq:twobo_arch}, in this work we found that a single dense layer is enough to obtain satisfactory results. This dense layer maps $N_{\xi i}$ inputs to one output, where $N_{\xi i}$ is the number of non-zero variables in $\vxi_i$. Note that we do not share the parameters among the functions $\rho_i$ with different $i$.
For comparison, we have tested the simplest MADE with a single dense layer, which still has more trainable parameters than the simplest TwoBo.
Another architecture we have compared with is the tensorized RNN introduced in~\cite{hibat-allah_variational_2021} for 2D classical spin systems, where we use $4$ memory units for a comparable number of trainable parameters, and a comparison with RNN at different sizes is shown in the Supplementary Material~\ref{append:rnn}. Details of the numerical experiments can be found in the Supplementary Material~\ref{append:details}.

\begin{figure}[htb]
\centering
\includegraphics[width=0.9\linewidth]{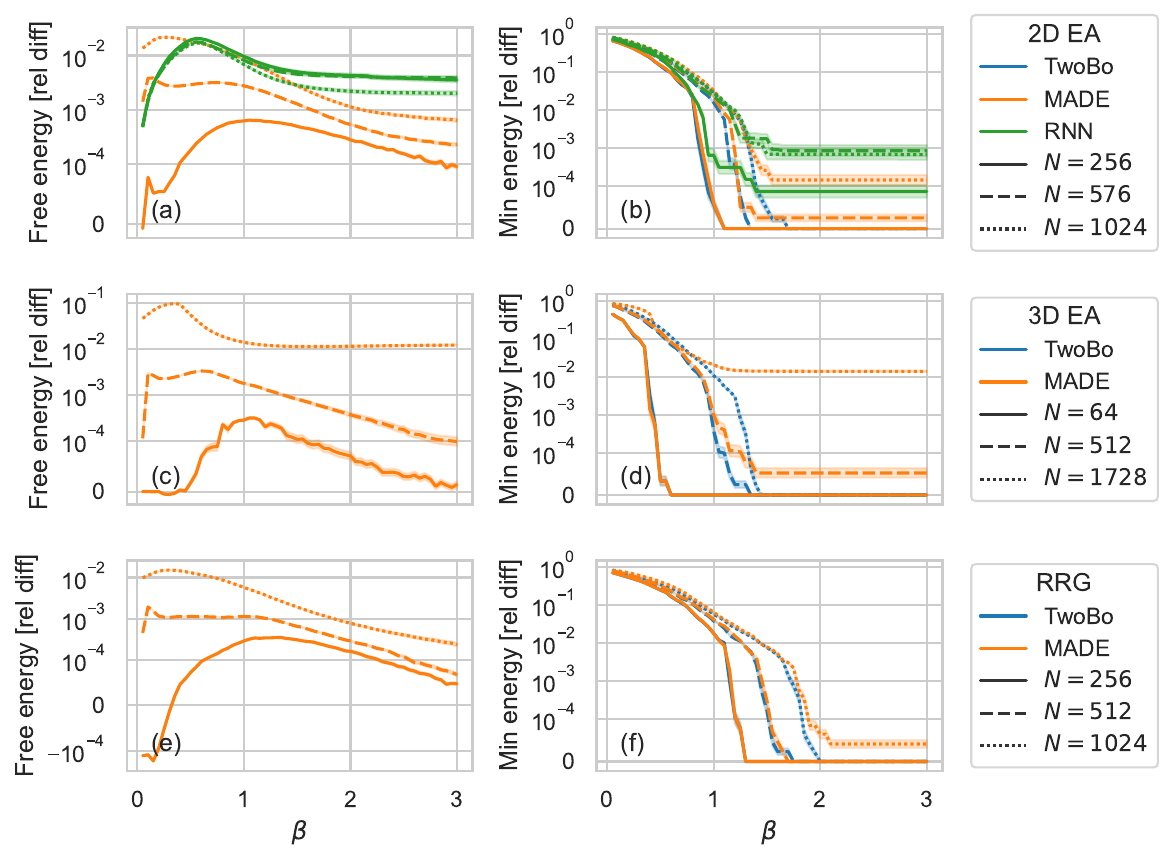}
\caption{
Variational free energy (left column) and minimum energy of sampled configurations (right column) at each $\beta$ on 2D EA (top row), 3D EA (middle row), and RRG (bottom row) models of different sizes. We compare the performances of TwoBo, MADE, and RNN (only in 2D).
The variational free energy is shown as the relative difference from TwoBo at the same $\beta$, and the minimum energy as the relative difference from the one found by TwoBo at $\beta = 3$.
The results are averaged over $10$ random instances of the Hamiltonian, and the error bars show the standard errors over the instances.
}
\label{fig:res_EA_RRG}
\end{figure}

In the left column of Fig.~\ref{fig:res_EA_RRG}, the free energies estimated by the three architectures are shown as a function of $\beta$, where the TwoBo result is taken as the reference for the relative difference.
We observe that TwoBo obtains the lowest free energy estimation on all three interaction graphs, except for smaller system sizes at the beginning of the annealing procedure where $\beta$ is low. This signifies a better approximation of the corresponding Boltzmann distribution.

Recent studies~\cite{ciarella2023machine, condmat7020038} have indicated the potential problem of mode collapse that could distort the free energy estimation in highly frustrated systems, where the variational distribution is trapped in local minima of free energy.
To corroborate the efficacy of TwoBo in capturing the high complexity of these systems, we have examined its ability to find ground state configurations, as shown in the right column of Fig.~\ref{fig:res_EA_RRG}.
MADE and RNN failed to find configurations with lower energy than TwoBo in large systems, indicating the difficulty in capturing the complexity of these probability distributions.

In addition to variational methods, we have used McGroundstate~\cite{CJMM22}, a recent max-cut solver for 2D and 3D lattices, to solve the EA instances and compared the results with TwoBo. McGroundstate solved all 2D instances with $N = 256, 576, 1024$ and 3D instances with $N = 64$ in provable optimality, as found by TwoBo, but it failed to provide results for larger 3D cases with $N \geq 512$. For the largest 3D cases, we have tried the traditional max-cut algorithm~\cite{burer02rank} implemented in MQLib~\cite{DunningEtAl2018}, which yielded energies higher than those obtained by TwoBo. Consequently, we chose the energy found by TwoBo as the reference for the relative difference in the minimum energy shown in the right column of Fig.~\ref{fig:res_EA_RRG}.

\begin{figure}[htb]
\centering
\includegraphics[width=0.9\linewidth]{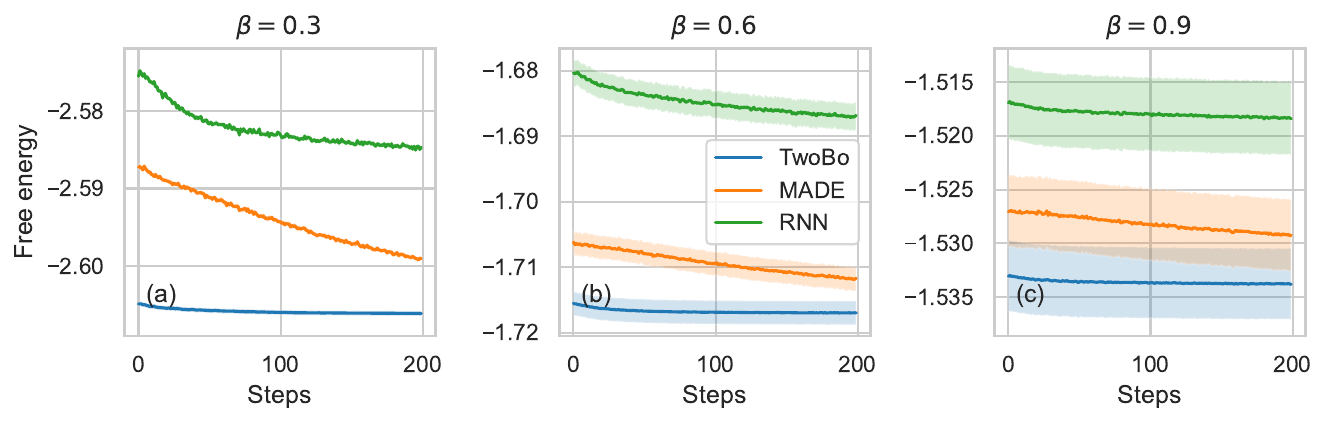}
\caption{
Convergence of variational free energy when training at fixed values of $\beta$ on 2D EA model with $N = 576$.
The results are averaged over $10$ random instances of the Hamiltonian, and the error bars show the standard errors over the instances.
}
\label{fig:converge}
\end{figure}

To investigate the convergence of the training procedure, Fig.~\ref{fig:converge} shows the free energy during the optimization steps at three different temperatures on a 2D EA model, from which we can observe that TwoBo always starts with a lower free energy compared to the other architectures. This advantage may stem from the fact that the parameters of the first layer are fixed to the Hamiltonian couplings, as well as the change of the skip connection coefficient with $\beta$, which leads the variational distribution to be closer to the target distribution at the beginning of the training procedure.
With significantly more optimization steps, MADE could eventually match the free energy obtained by TwoBo. In contrast, RNN converges to a higher free energy, indicating its insufficient expressivity or trainability in a regime of few trainable parameters.

\begin{figure}[htb]
\centering
\includegraphics[width=0.9\linewidth]{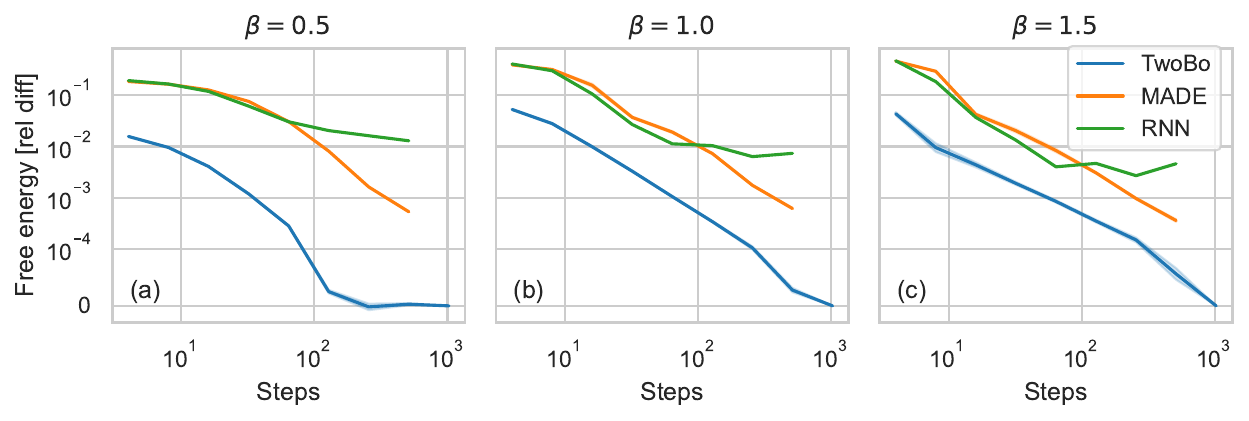}
\caption{
Variational free energy using different numbers of optimization steps at each $\beta$ on 2D EA model with $N = 576$, shown as the relative difference from TwoBo at the same $\beta$ using 1024 steps.
The results are averaged over $10$ random instances of the Hamiltonian, and the standard errors over the instances are too small to be visible.
}
\label{fig:converge_max_steps}
\end{figure}

To better assess this claim, Fig.~\ref{fig:converge_max_steps} displays the free energy as a function of the number of optimization steps in a whole temperature step. It is evident that as the number of steps increases, the free energy computed by TwoBo systematically decreases. MADE and RNN display a similar trend with almost an order of magnitude more steps, while RNN eventually reaches a plateau with a free energy higher than TwoBo. This comparison shows the abundant expressivity of TwoBo in capturing the target distribution and its ease of training.

\section*{Conclusion}

In this work, we presented a new autoregressive neural network architecture named TwoBo to sample and approximate the Boltzmann distribution of two-body interacting spin systems. It exploits the knowledge of the target Boltzmann distribution to determine part of its parameters, and systematically reduces the number of trainable parameters in sparse interacting systems without loss of expressivity. It obtains higher accuracy in approximating disordered and frustrated Boltzmann distributions with considerably fewer trainable parameters and faster convergence than previous ARNN architectures.

We achieve a reduction in the number of trainable parameters with respect to MADE in all cases considered. The most notable results are observed in the 2D lattice case, although the benefits decrease with increasing lattice dimensions. In contrast, in the long-range interaction case of RRG, while there is a reduction in parameters, the scaling remains comparable to that of MADE. Nevertheless, TwoBo still shows faster convergence and higher accuracy in these cases, thanks to the knowledge of the Boltzmann distribution.
Although the present work focuses on binary, two-body interacting variables, nothing prevents using a similar approach to derive ARNN architectures for multi-valued or continuous variables, such as the Potts model~\cite{wu1982potts} and the Kuramoto model~\cite{acebron2005kuramoto}, and with more than two-body interactions, such as the Baxter--Wu model~\cite{novotny1981critical}.
Beyond statistical physics, TwoBo has potential applications in a wide range of research areas that utilize Boltzmann distributions, including combinatorial optimization and inference problems, extending from mathematics to social sciences~\cite{doi:10.1126/science.1073287, doi:10.1080/00018732.2016.1211393, Biazzo2022, khandoker2023supplementing}. It also has promising uses in a family of energy-based models within the rapidly growing field of machine learning, particularly for tasks such as language understanding, image recognition, and other complex data interpretation activities~\cite{DBLP:conf/iclr/DengBOSR20, DBLP:conf/nips/DuLM20}.
We believe that the efficiency of TwoBo will help explore the computationally difficult regimes of these problems.

\section*{Acknowledgements}

The authors thank Simone Ciarella, Federico Ricci-Tersenghi, Jeanne Trinquier, Francesco Zamponi, Mohamed Hibat-Allah, and Juan Carrasquilla for valuable discussions.
Support from the Swiss National Science Foundation is acknowledged under Grant No.~200336.

\section*{Data availability statement}

The code for reproducing the results and the figures presented is openly available at the following URL: \url{https://github.com/cqsl/sparse-twobody-arnn} .

\section*{References}

\bibliographystyle{apalike}
\bibliography{ref}

\begin{thebibliography}{}

\bibitem[Acebr{\'o}n et~al., 2005]{acebron2005kuramoto}
Acebr{\'o}n, J.~A., Bonilla, L.~L., Vicente, C. J.~P., Ritort, F., and Spigler,
  R. (2005).
\newblock The {K}uramoto model: {A} simple paradigm for synchronization
  phenomena.
\newblock {\em Rev. Mod. Phys.}, 77(1):137.

\bibitem[Albert and Barab{\'a}si, 2002]{RevModPhys.74.47}
Albert, R. and Barab{\'a}si, A.-L. (2002).
\newblock Statistical mechanics of complex networks.
\newblock {\em Rev. Mod. Phys.}, 74(1):47--97.

\bibitem[Amey and Machta, 2018]{amey2018analysis}
Amey, C. and Machta, J. (2018).
\newblock Analysis and optimization of population annealing.
\newblock {\em Phys. Rev. E}, 97(3):033301.

\bibitem[Barab{\'a}si and Albert, 1999]{doi:10.1126/science.286.5439.509}
Barab{\'a}si, A.-L. and Albert, R. (1999).
\newblock Emergence of scaling in random networks.
\newblock {\em Science}, 286(5439):509--512.

\bibitem[Barrett et~al., 2022]{Barrett2022}
Barrett, T.~D., Malyshev, A., and Lvovsky, A.~I. (2022).
\newblock Autoregressive neural-network wavefunctions for ab initio quantum
  chemistry.
\newblock {\em Nature Mach. Intell.}, 4(4):351--358.

\bibitem[Bia{\l}as et~al., 2022]{BIALAS2022108502}
Bia{\l}as, P., Korcyl, P., and Stebel, T. (2022).
\newblock Hierarchical autoregressive neural networks for statistical systems.
\newblock {\em Comput. Phys. Commun.}, 281:108502.

\bibitem[Biazzo, 2023]{biazzo2023autoregressive}
Biazzo, I. (2023).
\newblock The autoregressive neural network architecture of the {B}oltzmann
  distribution of pairwise interacting spins systems.
\newblock {\em Commun. Phys.}, 6(1):296.

\bibitem[Biazzo et~al., 2022]{Biazzo2022}
Biazzo, I., Braunstein, A., Dall'Asta, L., and Mazza, F. (2022).
\newblock A bayesian generative neural network framework for epidemic inference
  problems.
\newblock {\em Sci. Rep.}, 12(1):19673.

\bibitem[Brown et~al., 2020]{NEURIPS2020_1457c0d6}
Brown, T., Mann, B., Ryder, N., Subbiah, M., Kaplan, J.~D., Dhariwal, P.,
  Neelakantan, A., Shyam, P., Sastry, G., Askell, A., Agarwal, S.,
  Herbert-Voss, A., Krueger, G., Henighan, T., Child, R., Ramesh, A., Ziegler,
  D., Wu, J., Winter, C., Hesse, C., Chen, M., Sigler, E., Litwin, M., Gray,
  S., Chess, B., Clark, J., Berner, C., McCandlish, S., Radford, A., Sutskever,
  I., and Amodei, D. (2020).
\newblock Language models are few-shot learners.
\newblock In {\em Advances in Neural Information Processing Systems},
  volume~33, pages 1877--1901. Curran Associates, Inc.

\bibitem[Burer et~al., 2002]{burer02rank}
Burer, S., Monteiro, R. D.~C., and Zhang, Y. (2002).
\newblock Rank-two relaxation heuristics for {MAX-CUT} and other binary
  quadratic programs.
\newblock {\em {SIAM} J. Optim.}, 12(2):503--521.

\bibitem[Cha et~al., 2021]{Cha_2022}
Cha, P., Ginsparg, P., Wu, F., Carrasquilla, J., McMahon, P.~L., and Kim, E.-A.
  (2021).
\newblock Attention-based quantum tomography.
\newblock {\em Mach. Learn.: Sci. Technol.}, 3(1):01LT01.

\bibitem[Charfreitag et~al., 2022]{CJMM22}
Charfreitag, J., J{\"u}nger, M., Mallach, S., and Mutzel, P. (2022).
\newblock {McSparse}: {E}xact solutions of sparse maximum cut and sparse
  unconstrained binary quadratic optimization problems.
\newblock In {\em 2022 Proceedings of the Symposium on Algorithm Engineering
  and Experiments ({ALENEX})}, pages 54--66.

\bibitem[Ciarella et~al., 2023]{ciarella2023machine}
Ciarella, S., Trinquier, J., Weigt, M., and Zamponi, F. (2023).
\newblock Machine-learning-assisted {M}onte {C}arlo fails at sampling
  computationally hard problems.
\newblock {\em Mach. Learn.: Sci. Technol.}, 4(1):010501.

\bibitem[Cipra, 2000]{cipra2000ising}
Cipra, B.~A. (2000).
\newblock The {I}sing model is {NP}-complete.
\newblock {\em {SIAM} News}, 33(6):1--3.

\bibitem[Cirauqui et~al., 2024]{cirauqui2024population}
Cirauqui, D., Garc{\'\i}a-March, M.~{\'A}., Saavedra, J. R.~M., Lewenstein, M.,
  and Grzybowski, P.~R. (2024).
\newblock Population annealing with topological defect driven nonlocal updates
  for spin systems with quenched disorder.
\newblock {\em Phys. Rev. B}, 109(14):144202.

\bibitem[Cocco et~al., 2018]{Cocco_2018}
Cocco, S., Feinauer, C., Figliuzzi, M., Monasson, R., and Weigt, M. (2018).
\newblock Inverse statistical physics of protein sequences: {A} key issues
  review.
\newblock {\em Rep. Prog. Phys.}, 81(3):032601.

\bibitem[Deng et~al., 2020]{DBLP:conf/iclr/DengBOSR20}
Deng, Y., Bakhtin, A., Ott, M., Szlam, A., and Ranzato, M. (2020).
\newblock Residual energy-based models for text generation.
\newblock In {\em 8th International Conference on Learning Representations,
  {ICLR} 2020}.

\bibitem[Dommers, 2017]{dommers2017metastability}
Dommers, S. (2017).
\newblock Metastability of the {I}sing model on random regular graphs at zero
  temperature.
\newblock {\em Probab. Theory Relat. Fields}, 167:305--324.

\bibitem[Du et~al., 2020]{DBLP:conf/nips/DuLM20}
Du, Y., Li, S., and Mordatch, I. (2020).
\newblock Compositional visual generation with energy based models.
\newblock In {\em Advances in Neural Information Processing Systems},
  volume~33.

\bibitem[Dunning et~al., 2018]{DunningEtAl2018}
Dunning, I., Gupta, S., and Silberholz, J. (2018).
\newblock What works best when? {A} systematic evaluation of heuristics for
  {Max-Cut} and {QUBO}.
\newblock {\em {INFORMS} J. Comput.}, 30(3).

\bibitem[Edwards and Anderson, 1975]{edwards1975theory}
Edwards, S.~F. and Anderson, P.~W. (1975).
\newblock Theory of spin glasses.
\newblock {\em J. Phys. F: Met. Phys.}, 5(5):965.

\bibitem[Germain et~al., 2015]{pmlr-v37-germain15}
Germain, M., Gregor, K., Murray, I., and Larochelle, H. (2015).
\newblock {MADE}: {M}asked autoencoder for distribution estimation.
\newblock In {\em Proceedings of the 32nd International Conference on Machine
  Learning ({ICML})}, volume~37, pages 881--889, Lille, France. PMLR.

\bibitem[He et~al., 2016]{He_2016_CVPR}
He, K., Zhang, X., Ren, S., and Sun, J. (2016).
\newblock Deep residual learning for image recognition.
\newblock In {\em Proceedings of the IEEE Conference on Computer Vision and
  Pattern Recognition ({CVPR})}.

\bibitem[Hibat-Allah et~al., 2020]{PhysRevResearch.2.023358}
Hibat-Allah, M., Ganahl, M., Hayward, L.~E., Melko, R.~G., and Carrasquilla, J.
  (2020).
\newblock Recurrent neural network wave functions.
\newblock {\em Phys. Rev. Res.}, 2(2):023358.

\bibitem[Hibat-Allah et~al., 2021]{hibat-allah_variational_2021}
Hibat-Allah, M., Inack, E.~M., Wiersema, R., Melko, R.~G., and Carrasquilla, J.
  (2021).
\newblock Variational neural annealing.
\newblock {\em Nature Mach. Intell.}, 3(11):952--961.

\bibitem[Inack et~al., 2022]{condmat7020038}
Inack, E.~M., Morawetz, S., and Melko, R.~G. (2022).
\newblock Neural annealing and visualization of autoregressive neural networks
  in the newman-moore model.
\newblock {\em Condensed Matter}, 7(2).

\bibitem[Khandoker et~al., 2023]{khandoker2023supplementing}
Khandoker, S.~A., Abedin, J.~M., and Hibat-Allah, M. (2023).
\newblock Supplementing recurrent neural networks with annealing to solve
  combinatorial optimization problems.
\newblock {\em Mach. Learn.: Sci. Technol.}, 4(1):015026.

\bibitem[Kingma and Ba, 2015]{kingma2014adam}
Kingma, D.~P. and Ba, J. (2015).
\newblock Adam: {A} method for stochastic optimization.
\newblock In {\em 3rd International Conference on Learning Representations,
  {ICLR} 2015}.

\bibitem[Liu et~al., 2021]{Liu_2021}
Liu, J.-G., Mao, L., Zhang, P., and Wang, L. (2021).
\newblock Solving quantum statistical mechanics with variational autoregressive
  networks and quantum circuits.
\newblock {\em Mach. Learn.: Sci. Technol.}, 2(2):025011.

\bibitem[Luo et~al., 2022]{10.1103/physrevlett.128.090501}
Luo, D., Chen, Z., Carrasquilla, J., and Clark, B.~K. (2022).
\newblock Autoregressive neural network for simulating open quantum systems via
  a probabilistic formulation.
\newblock {\em Phys. Rev. Lett.}, 128(9):090501.

\bibitem[McNaughton et~al., 2020]{PhysRevE.101.053312}
McNaughton, B., Milo{\v s}evi{\' c}, M.~V., Perali, A., and Pilati, S. (2020).
\newblock Boosting {M}onte {C}arlo simulations of spin glasses using
  autoregressive neural networks.
\newblock {\em Phys. Rev. E}, 101(5):053312.

\bibitem[Mezard and Montanari, 2009]{mezard2009information}
Mezard, M. and Montanari, A. (2009).
\newblock {\em Information, Physics, and Computation}.
\newblock Oxford University Press.

\bibitem[M{\'e}zard and Parisi, 2001]{mezard2001bethe}
M{\'e}zard, M. and Parisi, G. (2001).
\newblock The {B}ethe lattice spin glass revisited.
\newblock {\em Eur. Phys. J. B}, 20:217--233.

\bibitem[Mezard et~al., 1986]{10.1142/0271}
Mezard, M., Parisi, G., and Virasoro, M. (1986).
\newblock {\em Spin Glass Theory and Beyond}.
\newblock World Scientific.

\bibitem[M{\'e}zard et~al., 2002]{doi:10.1126/science.1073287}
M{\'e}zard, M., Parisi, G., and Zecchina, R. (2002).
\newblock Analytic and algorithmic solution of random satisfiability problems.
\newblock {\em Science}, 297(5582):812--815.

\bibitem[Nicoli et~al., 2020]{10.1103/physreve.101.023304}
Nicoli, K.~A., Nakajima, S., Strodthoff, N., Samek, W., M{\"u}ller, K.-R., and
  Kessel, P. (2020).
\newblock Asymptotically unbiased estimation of physical observables with
  neural samplers.
\newblock {\em Phys. Rev. E}, 101(2):023304.

\bibitem[Novotny and Landau, 1981]{novotny1981critical}
Novotny, M.~A. and Landau, D.~P. (1981).
\newblock Critical behavior of the {B}axter--{W}u model with quenched
  impurities.
\newblock {\em Phys. Rev. B}, 24(3):1468.

\bibitem[Pan et~al., 2021]{PhysRevE.103.012103}
Pan, F., Zhou, P., Zhou, H.-J., and Zhang, P. (2021).
\newblock Solving statistical mechanics on sparse graphs with feedback-set
  variational autoregressive networks.
\newblock {\em Phys. Rev. E}, 103(1):012103.

\bibitem[Picco et~al., 2001]{picco2001chaotic}
Picco, M., Ricci-Tersenghi, F., and Ritort, F. (2001).
\newblock Chaotic, memory, and cooling rate effects in spin glasses:
  {E}valuation of the {E}dwards--{A}nderson model.
\newblock {\em Phys. Rev. B}, 63(17):174412.

\bibitem[Sharir et~al., 2020]{PhysRevLett.124.020503}
Sharir, O., Levine, Y., Wies, N., Carleo, G., and Shashua, A. (2020).
\newblock Deep autoregressive models for the efficient variational simulation
  of many-body quantum systems.
\newblock {\em Phys. Rev. Lett.}, 124(2):020503.

\bibitem[Stein and Newman, 2013]{stein2013spin}
Stein, D.~L. and Newman, C.~M. (2013).
\newblock {\em Spin Glasses and Complexity}, volume~4.
\newblock Princeton University Press.

\bibitem[Wang and Davis, 2020]{PhysRevA.102.062413}
Wang, Z. and Davis, E.~J. (2020).
\newblock Calculating {R}{\'e}nyi entropies with neural autoregressive quantum
  states.
\newblock {\em Phys. Rev. A}, 102(6):062413.

\bibitem[Wu et~al., 2021]{PhysRevResearch.3.L042024}
Wu, D., Rossi, R., and Carleo, G. (2021).
\newblock Unbiased {M}onte {C}arlo cluster updates with autoregressive neural
  networks.
\newblock {\em Phys. Rev. Res.}, 3(4):L042024.

\bibitem[Wu et~al., 2023]{PhysRevResearch.5.L032001}
Wu, D., Rossi, R., Vicentini, F., and Carleo, G. (2023).
\newblock From tensor-network quantum states to tensorial recurrent neural
  networks.
\newblock {\em Phys. Rev. Res.}, 5(3):L032001.

\bibitem[Wu et~al., 2019]{PhysRevLett.122.080602}
Wu, D., Wang, L., and Zhang, P. (2019).
\newblock Solving statistical mechanics using variational autoregressive
  networks.
\newblock {\em Phys. Rev. Lett.}, 122(8):080602.

\bibitem[Wu, 1982]{wu1982potts}
Wu, F.-Y. (1982).
\newblock The {P}otts model.
\newblock {\em Rev. Mod. Phys.}, 54(1):235.

\bibitem[Zdeborov{\'a} and Krzakala, 2016]{doi:10.1080/00018732.2016.1211393}
Zdeborov{\'a}, L. and Krzakala, F. (2016).
\newblock Statistical physics of inference: {T}hresholds and algorithms.
\newblock {\em Adv. Phys.}, 65(5):453--552.

\end{thebibliography}

\clearpage

\vspace*{-2ex}
\begin{center}
\textbf{\large Supplementary Material}
\end{center}

\setcounter{section}{0}
\setcounter{equation}{0}
\setcounter{figure}{0}
\setcounter{table}{0}

\renewcommand{\thesection}{\Alph{section}}
\renewcommand{\theequation}{S\arabic{equation}}
\renewcommand{\thefigure}{S\arabic{figure}}
\renewcommand{\thetable}{S\arabic{table}}

\section{Scaling of number of trainable parameters} \label{append:param}

\begin{figure}[htb]
\centering
\includegraphics[width=0.9\linewidth]{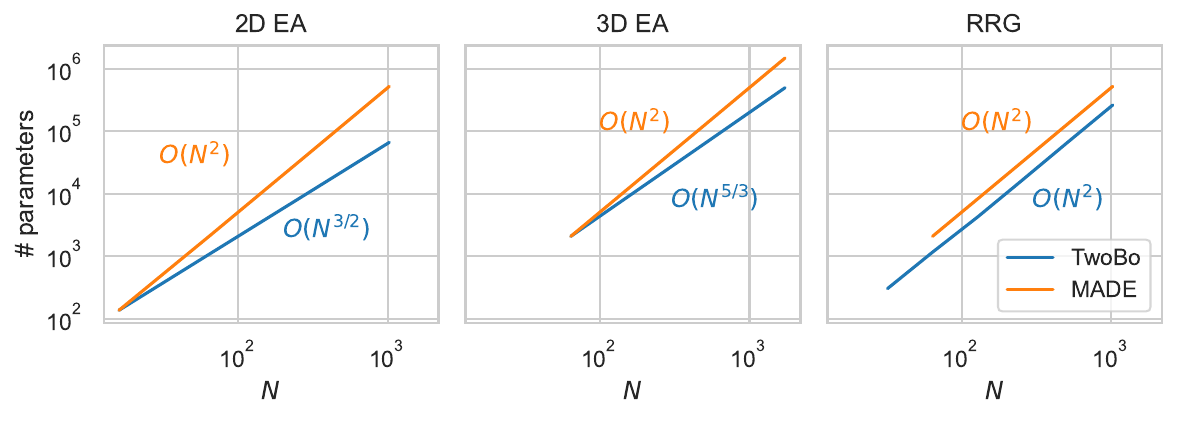}
\caption{
Number of trainable parameters in TwoBo and MADE on 2D EA, 3D EA, and RRG models of various system sizes $N$.
For RRG, the number of trainable parameters in TwoBo is averaged over $10$ random instances.
}
\label{fig:compare_param}
\end{figure}

As shown in Fig.~\ref{fig:compare_param}, on 2D and 3D EA models, the number of trainable parameters in TwoBo scales polynomially slower than that in MADE. On RRG model without a regular grid geometry, TwoBo has the same scaling as MADE, and it only needs half as many parameters.

\section{Annealing $\beta$ in the skip connection coefficient} \label{append:beta}

\begin{figure}[htb]
\centering
\includegraphics[width=0.5\linewidth]{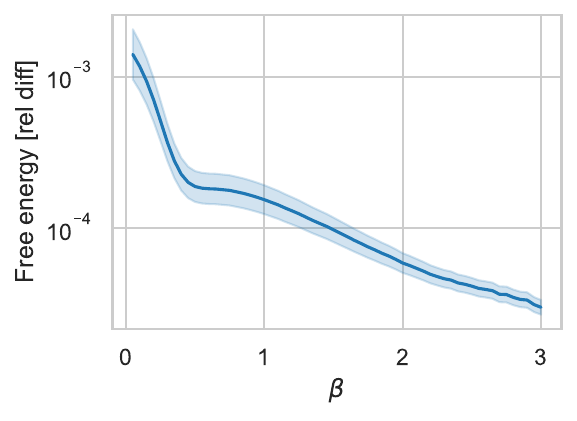}
\caption{
Variational free energy when fixing $\beta = 1$ in the skip connection coefficient, shown as the relative difference from updating it during the annealing procedure, on 2D EA model with $N = 576$.
The results are averaged over $10$ random instances of the Hamiltonian, and the error bars show the standard errors over the instances.
}
\label{fig:compare_beta}
\end{figure}

When optimizing the TwoBo architecture, we not only gradually change the $\beta$ used to compute the variational free energy, but also change the $\beta$ in the skip connection coefficient in Eq.~\eqref{eq:twobo_arch} to the same value at each annealing step. If we do not update the $\beta$ in the skip connection coefficient, but fix it to $1$, the resulting variational free energy will be higher especially at the beginning of the annealing procedure, as shown in Fig.~\ref{fig:compare_beta}.

\section{Comparison with RNN at different sizes} \label{append:rnn}

\begin{figure}[htb]
\centering
\includegraphics[width=0.9\linewidth]{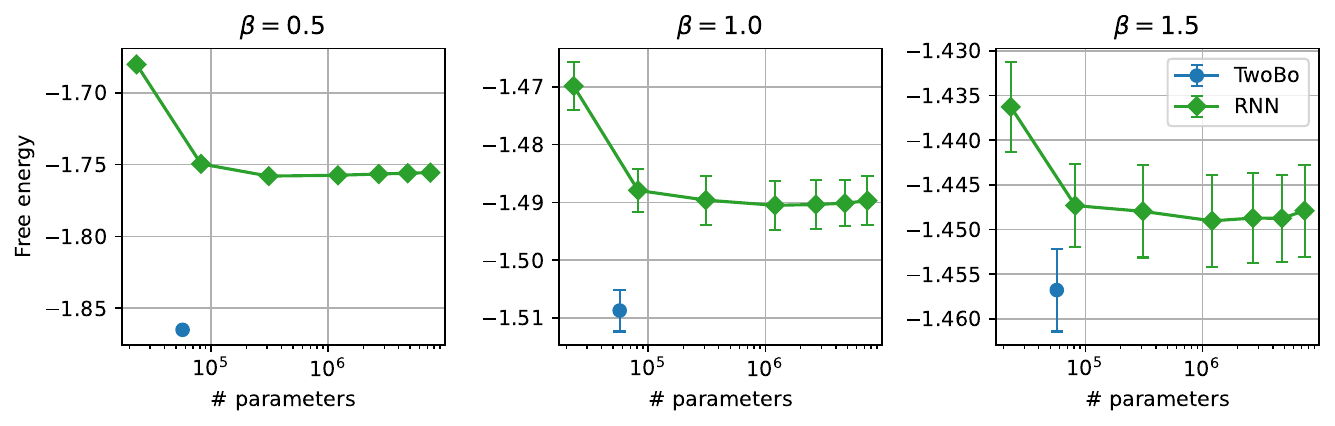}
\caption{
Variational free energy of TwoBo compared to RNN with different numbers of memory units ($2, 4, 8, 16, 24, 32, 40$), on 2D EA model with $N = 576$.
The results are averaged over $10$ random instances of the Hamiltonian, and the error bars show the standard errors over the instances.
}
\label{fig:compare_rnn}
\end{figure}

Fig.~\ref{fig:compare_rnn} shows that TwoBo gives lower free energy than RNN with a comparable number of trainable parameters, and the RNN results do not qualitatively change even if we increase the number of trainable parameters by orders of magnitude. When separately examining the estimations of the energy and the entropy, we can see that the high free energy of RNN is mainly because the entropy is low, which indicates that it is more impacted by mode collapse compared to TwoBo.

\section{Details of numerical experiments} \label{append:details}

In TwoBo and MADE, the only dense layer is initialized using Lecun normal initialization (the default to initialize dense layers in JAX) multiplied by a small factor of $0.01$, so that the output values before the last sigmoid function are small, and after the sigmoid function the probability distribution is almost uniform at the beginning of training, which helps alleviate mode collapse.

The implementation of tensorized RNN is taken from \url{https://github.com/mhibatallah/RNNWavefunctions}. We only used their neural network architecture but not their annealing procedure in training, because our annealing procedure aims to produce a converged free energy at each step of $\beta$, while theirs only aims to produce a ground state energy at zero temperature.

At the beginning of training, we run $500$ optimization steps as warm-up with the initial $\beta = 0.05$. Then for each $\beta$ in steps of $0.05$ until $\beta = 3$, we run $200$ optimization steps. Note that the value of $\beta$ is unchanged during those optimization steps. After each step of $\beta$, the trained neural network is used as the initialization in the next $\beta$. The optimizer we use is Adam~\cite{kingma2014adam} with learning rate $10^{-3}$, and at each step we take batch size $1024$.

The max-cut solver McGroundstate provides public access through their website. They put a time limit of $1800$ seconds for each spin glass instance, and they failed to solve the 3D EA models with $N = 512$ and $1728$ within the time limit. We have run MQLib on our local machine with a time limit of $24$ hours. It outputs the energy whenever it finds a lower one, and it stopped improving the energy after the first hour.

\end{document}